\documentclass{aa}  
\usepackage{graphicx}
\usepackage{txfonts}
\usepackage{natbib}
\bibpunct{(}{)}{;}{a}{}{,} 
\begin{document}
   \title{New CO detections of lensed submillimeter galaxies in A2218: Probing molecular gas in the LIRG regime at high redshift \thanks{Based on observations obtained at the IRAM Plateau de Bure Interferometer (PdBI).  IRAM is funded by the Centre National de la Recherche Scientifique (France), the Max-Planck Gesellschaft (Germany), and the Instituto Geografico Nacional (Spain).}}

   \titlerunning{New CO detection of lensed SMGs}

   \author{K.K. Knudsen
          \inst{1}
          \and
          R. Neri\inst{2}
          \and 
          J.-P. Kneib\inst{3}
          \and
          P.P. van der Werf\inst{4}
          }

   \offprints{K.K. Knudsen}

   \institute{Argelander-Institut f\"ur Astronomie, Auf dem H\"ugel 71, D-53123 Bonn, Germany \\
              \email{knudsen@astro.uni-bonn.de}
         \and
              Institut de Radio Astronomie Millim\'etrique (IRAM), 300 Rue de la Piscine, Domaine Universitaire de Grenoble, St. Martin d'H\`eres F-38406, France 
              \and Laboratoire d'Astrophysique de Marseille, OAMP, Universit\'e Aix-Marseille \& CNRS, 38 rue F.~Joliot-Curie, 13388 Marseille Cedex 13, France
              \and Leiden Observatory, Leiden University, P.O.Box 9513, NL-2300 RA Leiden, Netherlands
             }

   \date{}

 
  \abstract
   {Submillimeter galaxies (SMGs) are distant, dusty galaxies undergoing 
star formation at prodigious rates.  Recently there has been major progress
in understanding the nature of the bright SMGs (i.e. $S_{\rm 850\mu m} > 5$
mJy).  The samples for the fainter SMGs are small and are currently in a
phase of being built up through identification studies. } 
  {We study the molecular gas content in two SMGs, SMMJ163555 and SMMJ163541,
at redshifts $z=1.034$ and $z=3.187$ with un-lensed submillimeter fluxes of
0.4 mJy and 6.0 mJy.  Both SMGs are gravitationally lensed by the foreground
cluster Abell~2218. }
   {IRAM Plateau de Bure Interferometry observations at 3mm were obtained for
the lines CO(2-1) for SMMJ163555 and CO(3-2) for SMMJ163541.  Additionally we
obtained CO(4-3) for the candidate $z=4.048$ SMMJ163556 with an unlensed
submm flux of 2.7\,mJy. }
   {CO(2-1) was detected for SMMJ163555 at redshift 1.0313 with an integrated
line intensity of $1.2\pm0.2$ Jy km/s and a line width of $410\pm120$ km/s.
From this a gas mass of $1.6\times10^{9}$\,M$_\odot$ is derived and a star
formation efficiency of 440\,L$_\odot$/M$_\odot$ is estimated.  CO(3-2) was
detected for SMMJ163541 at redshift 3.1824, possibly with a second component
at redshift 3.1883, with an integrated line intensity of $1.0\pm0.1$ Jy km/s
and a line width of $280\pm50$ km/s.  From this a gas mass of
$2.2\times10^{10}$\,M$_\odot$ is derived and a star formation efficiency of
1000\,L$_\odot$/M$_\odot$ is estimated.  For SMMJ163556 the CO(4-3) is
undetected within the redshift range $4.035-4.082$ down to a sensitivity of
0.15\,Jy\,km/s.  }
   {Our CO line observations confirm the optical redshifts for SMMJ163555 and
SMMJ163541.  The CO line luminosity $L'_{\rm CO}$ for both galaxies is
consistent with the $L_{\rm FIR}-L'_{\rm CO}$ relation.  
SMMJ163555 has the lowest far-infrared luminosity of all SMGs with a known
redshift and is one of the few high redshift LIRGs whose properties can be
estimated prior to ALMA.}


   \maketitle
%

\section{Introduction}

CO observations of high redshift galaxies are instrumental to assess the
quantity of molecular gas available for star formation.  The submillimeter
galaxies \citep[SMGs; see][for review]{blain02} are dusty galaxies undergoing
star formation at prodigious rates, $> 1000$\,M$_\odot$yr$^{-1}$ with a
redshift distribution peaking at $z\sim2.5$ \citep{chapman05}.  Previous CO
results of SMGs have shown that SMGs can harbour molecular gas reservoirs
on
the order of $10^{10}$\,M$_\odot$, enabling the massive starburst to build-up
stars during 10-100 Myrs.  It is thus likely that SMGs are the
progenitors of the nearby massive ellipticals 
\citep[e.g.][]{chapman05,greve05,lapi06,tacconi06,nesvadba07}. 

Hitherto the CO results of SMGs have primarily been obtained for SMGs with
intrinsic fluxes $S_{\rm 850\mu m} > 2$\,mJy
\citep{frayer98,frayer99,neri03,genzel03,downes03,tacconi06,tacconi08}.  Of
the five known SMGs with lensing corrected $S_{\rm 850\mu m}< 1$\,mJy
\citep{kneib04a,borys04,knudsen06,knudsen08a}, only the triply-imaged galaxy
SMMJ16359+6612 has previously been detected in CO
\citep{sheth04,kneib05,weiss05}. 
Through statistical studies it is clear that other high redshift
galaxies populations, such distant red galaxies, extremely red objects and
BzK galaxies, have an average 850$\mu$m flux $\leq 1$\,mJy
\citep[e.g.][]{webb04,knudsen05,daddi05}, and thus there will be an overlap
between these and the faint SMGs.  However, none of these have individual
detections at these low flux density levels due to the SCUBA 850$\mu$m
confusion limit making it difficult to establish the exact link between the
faint SMGs and other galaxy populations.

In this paper we report the IRAM Plateau de Bure Interferometer (PdBI) 
observations of two strongly lensed SMGs, both present in the galaxy 
cluster field Abell~2218 \citep{knudsen06}.  
SMM\,J163555.2+661150 (henceforth SMMJ163555) has an optical counterpart at
redshift $z=1.034$, which is also known as galaxy \#289 in the notation of
\citet{pello92}.  It has an unlensed 850$\mu$m flux of 0.4 mJy and is one of
the lowest redshift SMGs \citep[cf.\ SMM\,J02396-0134;][]{smail02}, making it
the SMG with the lowest far-infrared luminosity known so far. 
SMM\,J163541.2+661144 (henceforth SMMJ163541) has been identified with a
galaxy at redshift $z=3.187$; optical spectroscopy is also presented in
this paper.  With
its unlensed 850$\mu$m flux of 6.0\,mJy it belongs to the bright SMGs as
found typically in the blank field surveys. 
SMMJ163555.5+661300 (henceforth SMMJ163556) has an unlensed 850$\mu$m flux of
2.7\,mJy and has a candidate optical counterpart at redshift 
$z=4.048$ \citep{knudsen08b} placing it possibly among the highest redshift
SMGs known.  

Throughout the paper we assume a cosmology 
with $H_0 = 70$\,km/s/Mpc, $\Omega_m = 0.3$ and $\Omega_\Lambda = 0.7$. 


\section{Observations}
\label{sect:obs}

\subsection{Optical spectroscopy}

\begin{figure}
\includegraphics[angle=270,width=8cm]{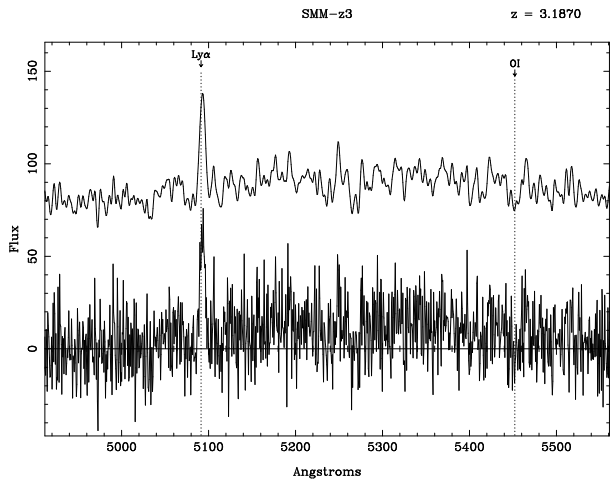}
\caption[]{The optical spectrum of SMMJ163541 is shown, where the dotted
lines show the Ly$\alpha$ and OI features.  
On the bottom, we show the original spectrum and on the top the
spectrum
is smoothed using a $\sigma=3$ pixel Gaussian and shifted in
the vertical direction for clarity.
\label{fig:z3-spec}}
\end{figure}

On 2003 June 30 and July 1, we conducted deep multislit spectroscopy with a
low resolution imaging spectrograph (LRIS; Oke et al. 1995) at the 10m
W.M.~Keck telescope at Mauna Kea, Hawaii, of sources lying
in the field of the rich cluster A2218 (see Kneib et al., 2004). The two
nights had reasonable seeing, $\sim$0.8 arcsec, but were not fully photometric
(with some cirrus), nevertheless we obtained a crude flux calibration of our
observations using Feige 67 and 110 as spectrophotometric standard stars.  We
observed SMM163541 for a total of 2.9 hrs using the 600/4000 blue
grism and the 400/8500 red gratings offering a spectral dispersion of 0.63
\AA\,pixel$^{−1}$ in the blue and 1.86 \AA\,pixel$^{-1}$ in the red, respectively.  The
spectrum of SMM163541 shows a strong Ly$\alpha$ emission and a possible OI
metal line (Fig.~\ref{fig:z3-spec}). The redshift derived is
$z=3.187\pm0.001$ based on Ly-alpha emission. 

Optical spectroscopy for SMMJ163555 and SMMJ163556 have been presented
elsewhere \citep{pello92,knudsen08b}. 

\subsection{IRAM Plateau de Bure}

\begin{figure*}
\begin{center}
\includegraphics[angle=270, width=13cm]{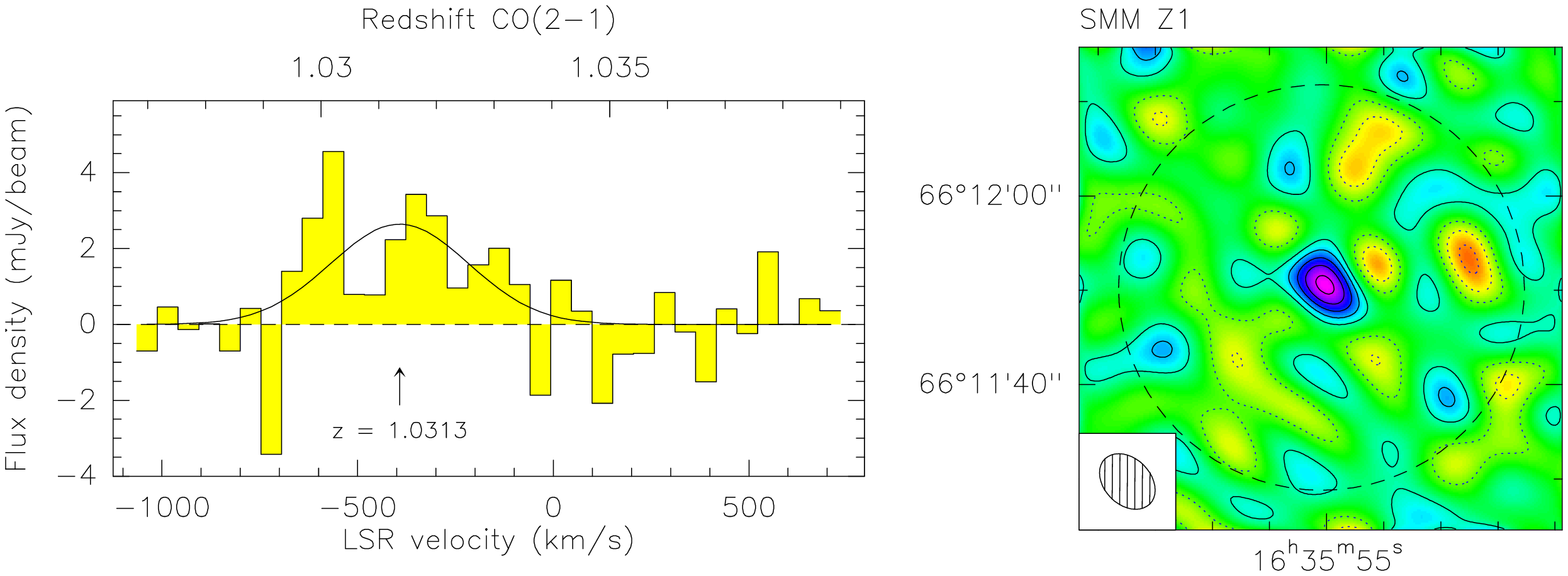} \\
\includegraphics[angle=270, width=13cm]{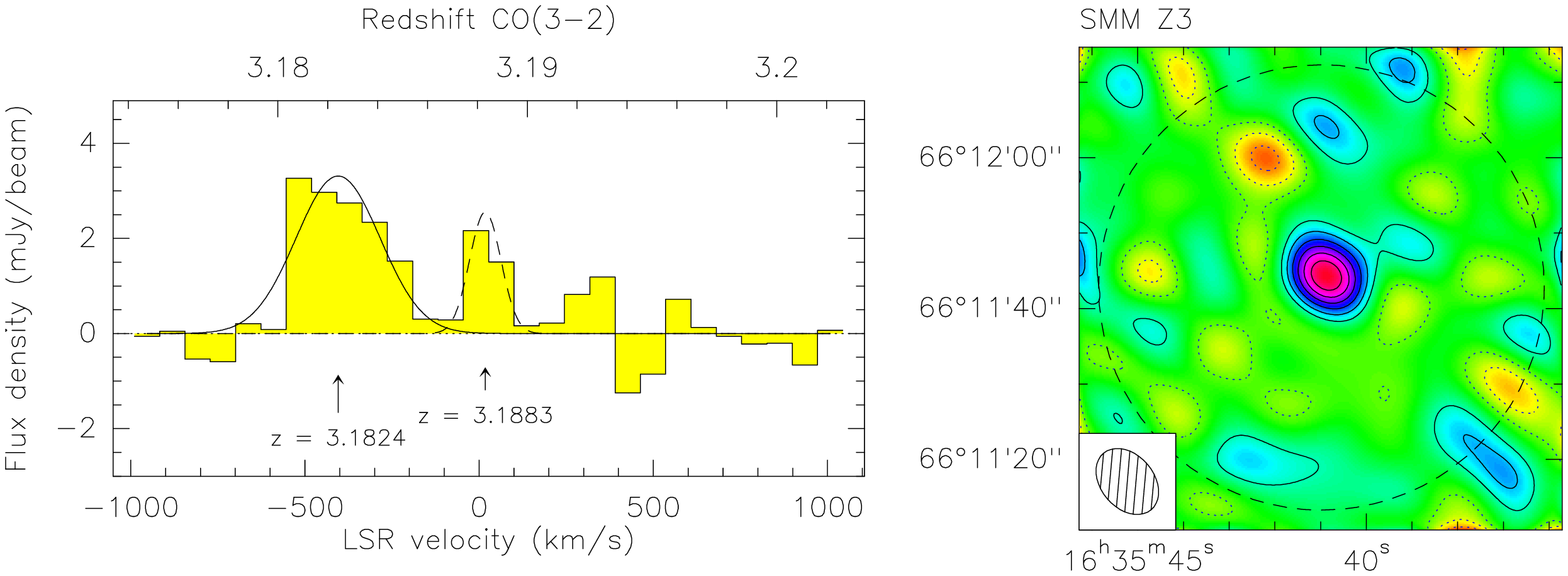} \\
\includegraphics[angle=270, width=13cm]{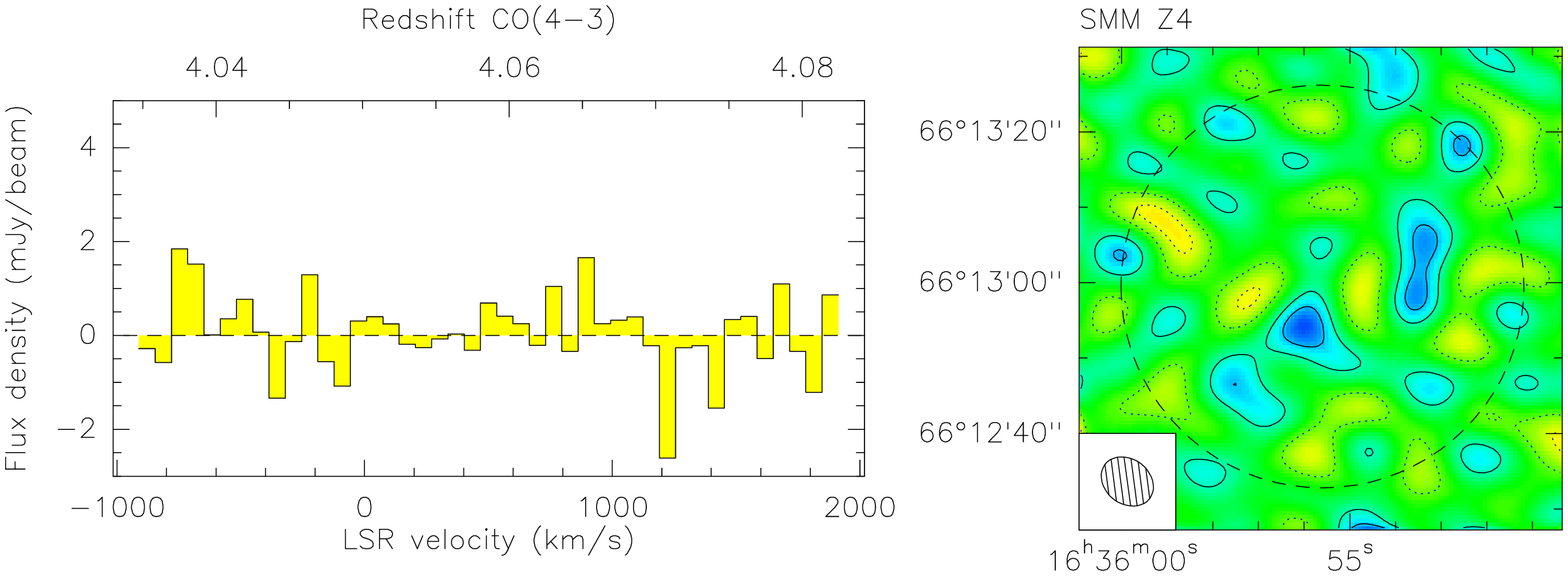}
\end{center}
\caption[]{
The resulting spectra and maps from the IRAM CO observations. 
Top panel:  The CO(2$\to$1) detection of SMMJ163555 ($z=1.0313$).
Middle panel:  The CO(3$\to$2) detection of SMMJ163541 ($z=3.1824$). 
Bottom panel:  The extracted spectrum at the position of SMMJ163556.
The velocity intervals used for producing the maps on the right side
are: SMMJ163555: -695 to -75 km/s, SMMJ163541: -265 to -555 km/s, and
SMMJ163556: the whole velocity range. 
The contours in the maps are in steps of $1\sigma$, solid and dashed 
show positive and negative contours, respectively. 
The large dashed circle indicates the primary beam.
\label{fig:res}}
\end{figure*}

Observations of SMMJ163555, SMMJ163541, and SMMJ163556 were carried out 
during May, June, July, August and September 2005 with the 
IRAM PdBI consisting of six 15m diameter antennas in the compact 
D configuration.  The correlator was configured for the CO line and 
continuum observations to simultaneously cover 580 MHz in the 3 and 
1.3mm bands.  
In the case of SMMJ163556, the frequency setting was re-adjusted during 
the observations to add more observations to a tentative detection. 
Consequently, the sensitivity varies across the final spectrum.  
The total integration time was 15.9 hours, 14.3 hours, and 17.3 hours 
for SMMJ163555, SMMJ163541, and SMMJ163556, respectively.   
All sources were observed in good observing conditions, where the typical
system temperatures at 3mm were $100-250$\,K. 
The data were reduced and analysed using the IRAM {\sc GILDAS} software. 
For passband calibration bright quasars were used. 
The phase and amplitude variations within each track were 
calibrated out by interleaving reference observations of the calibrator 
source 1637+574 
every 20 minutes.  The overall flux scale for each epoch was set 
on MWC\,349. 
Finally, naturally weighted data cubes were created, and the final beam 
in the maps at 3mm were $6.8''\times 4.9''$, $9.8''\times7.2''$, and 
$7.4''\times5.9''$ for SMMJ163555, SMMJ163541, and SMMJ163556, respectively.


\section{Results}
\label{sect:res}

\begin{table*}
\caption[]{Observational results for SMMJ163555, SMMJ163541, SMMJ163556 
\label{tab:res}}
\begin{tabular}{ccccccccc}
\hline
\hline
 & & & R.A. & Decl. & $\sigma_{\rm pos}$$^\diamond$ & $I_{\rm CO}$ & Flux$^\dagger$ & Line Width \\
Source & Transition & $z_{\rm CO}$ & (J2000.0) & (J2000.0) & [$''$] & [Jy km/s] & [mJy] & [km/s] \\ 
\hline
SMMJ163555 & CO(2-1) & $1.0313\pm0.0012$ & 16 35 55.05& +66 11 50.7 &
$\sim0.5$ & $1.2\pm0.2$ & $2.6\pm0.7$ & $410\pm120$ \\
 & 3mm &     &     &    &  &     & $<1.4$ &     \\
SMMJ163541 & CO(3-2) & $3.1824\pm0.0017$ & 16 35 40.86 & +66 11 44.4 & $\sim0.4$ & $1.0\pm0.1$ & $3.3\pm0.4$ & $284\pm50$ \\
           & CO(3-2) & $3.1883\pm0.0006$ & 16 35 41.33 & +66 11 46.4 & $\sim0.8$ & $0.3\pm0.1$ & $2.6\pm0.7$ & $106\pm44$ \\
 & 3mm &     &     &   &   &     & $<0.6$ &     \\
SMMJ163556 & CO(4-3) & $4.035-4.082$$^{\ddagger}$ &            &     &              &  $<0.45^\star$ & &  \\
 & 3mm &     &     &  &    &     & $<0.44$ &     \\
\hline
\end{tabular}
\\ 
$^\dagger$ For line emission this refers to the peak flux, for continuum this
refers to the flux density. \\
$^\ddagger$ The corresponding redshift ranged in which the CO(4-3) transition
would have been detectable. \\
$^\star$ $3\sigma$ upper limit estimated assuming a line width of 400 km/s.\\
$^\diamond$ $\sigma_{\rm pos}$ refers to the uncertainty on the position. 
\end{table*}

\begin{table*}
\begin{center}
\caption[]{Derived properties for SMMJ163555 and SMMJ163541.
\label{tab:prop} }
\begin{tabular}{lcccccc}
\hline
\hline
& $D_A$ [Gpc] & $\mu$\,$^\mathrm{a}$ & $D=1''$ [kpc] & $L{\rm '_{CO}}$ [K km/s pc$^2$] & $M_{\rm gas}$ [$M_\odot$] & $L_{\rm FIR}$ [$L_\odot$] \\
\hline
SMMJ163555 & 1.66 & 7.1 & 1.1 & $2.3\times10^{9}$   & $1.8\times10^{9}$  &
$4.5\times10^{11}$  \\
SMMJ163541 & 1.57 & 1.7 & 4.4 & $2.8\times10^{10}$  & $2.2\times10^{10}$ &
$1.5\times10^{13}$ \\
\ \ \ \ \ --\ \  two comp. & & &     & $3.6\times10^{10}$ & 
$2.9\times10^{10}$ &  \\
SMMJ163556 & 1.42 & 4.2 & 1.6 & $<0.4\times10^{10}$$^\mathrm{b}$ & $<0.3\times10^{10}$ &
$8.0\times10^{12}$ \\
\hline
\end{tabular}
\begin{list}{}{}
\item[$^{\mathrm{a}}$]Gravitational lensing magnification
\item[$^{\mathrm{b}}$]Based on the upper limit from table \ref{tab:res}
assuming a line width of 400 km/s.
\end{list}
\end{center}
\end{table*}

\begin{figure}
\begin{center}
\includegraphics[width=7cm]{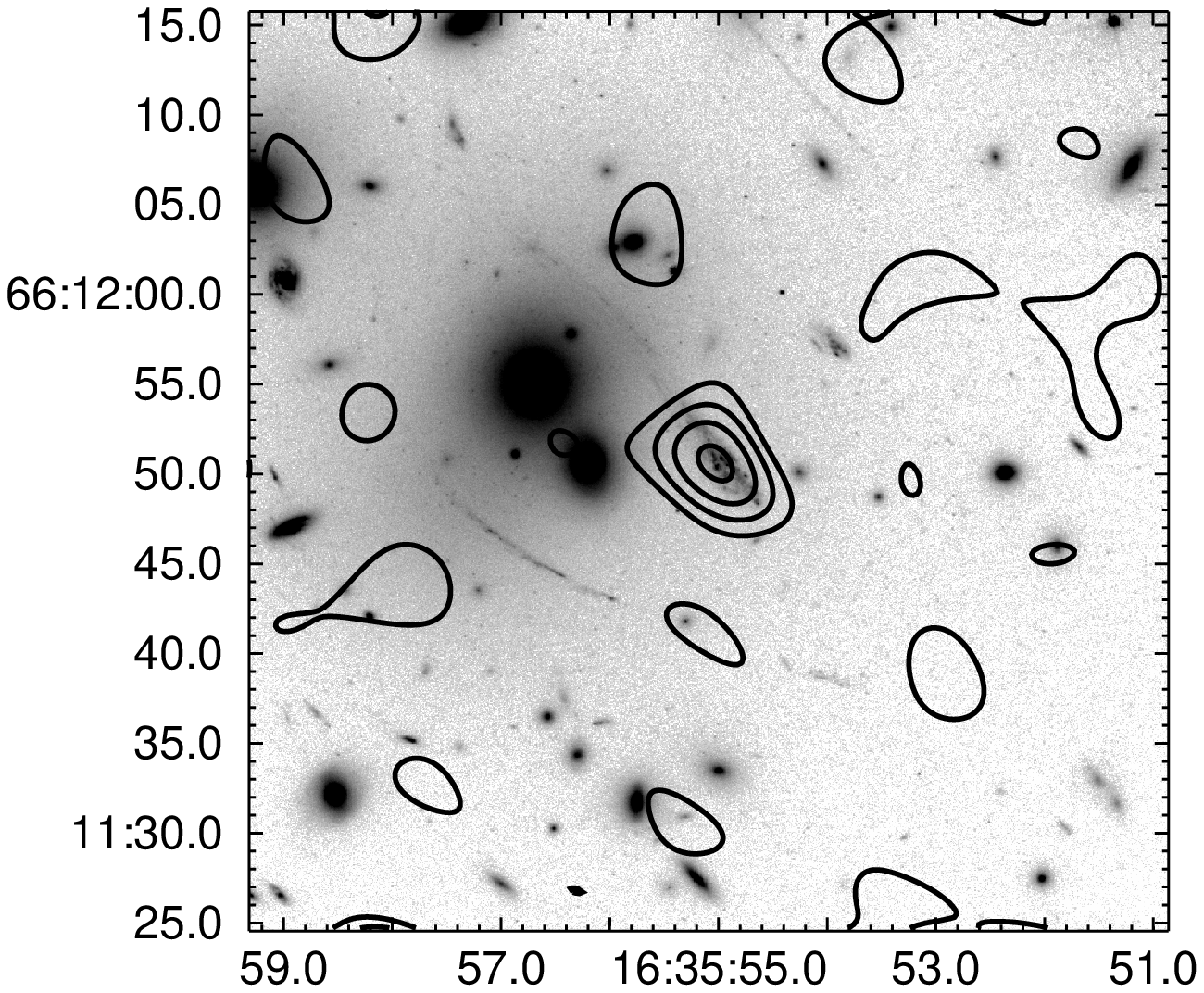} \\ \vskip0.7cm
\includegraphics[width=7cm]{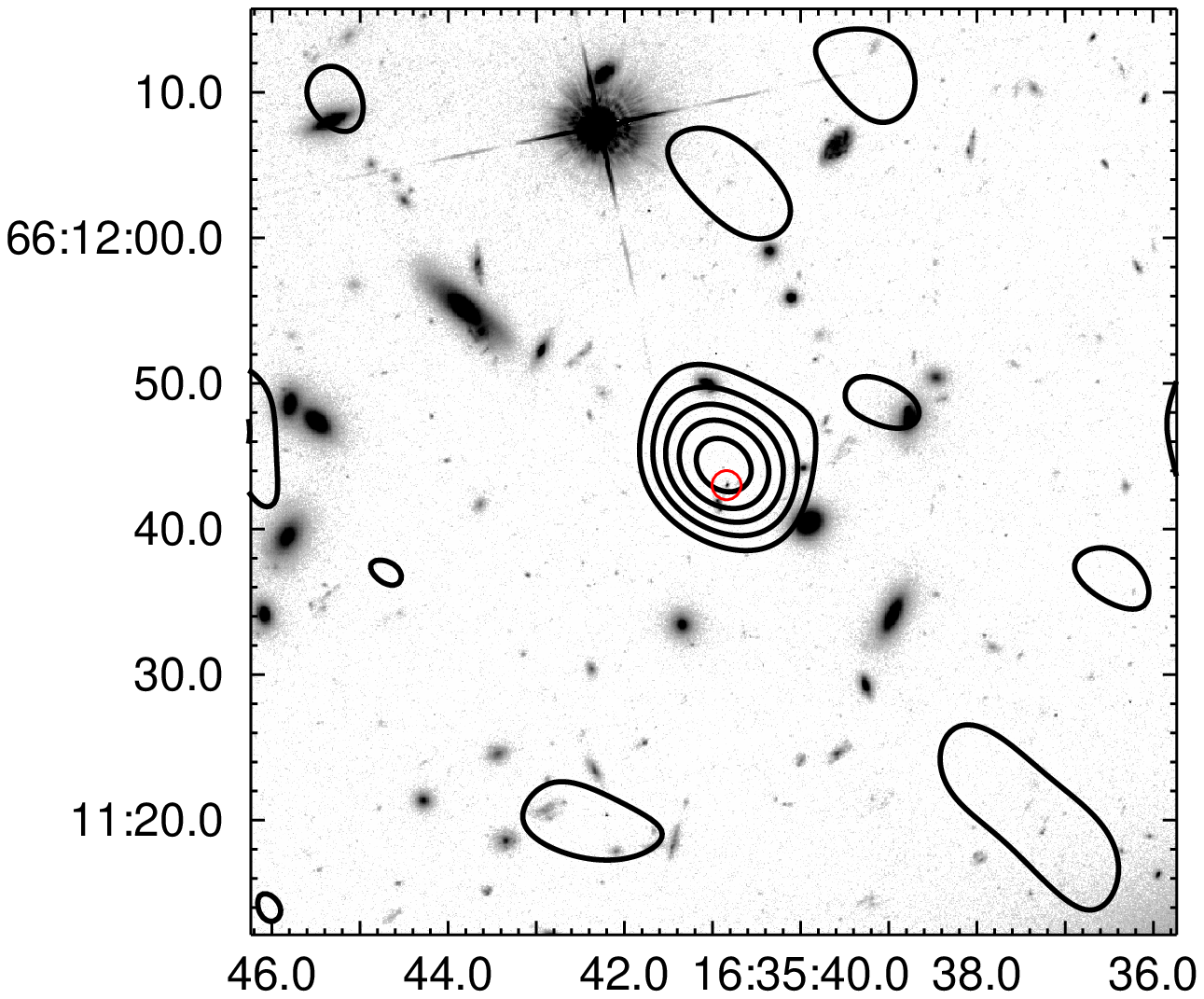} \\ \vskip0.7cm
\includegraphics[width=7cm]{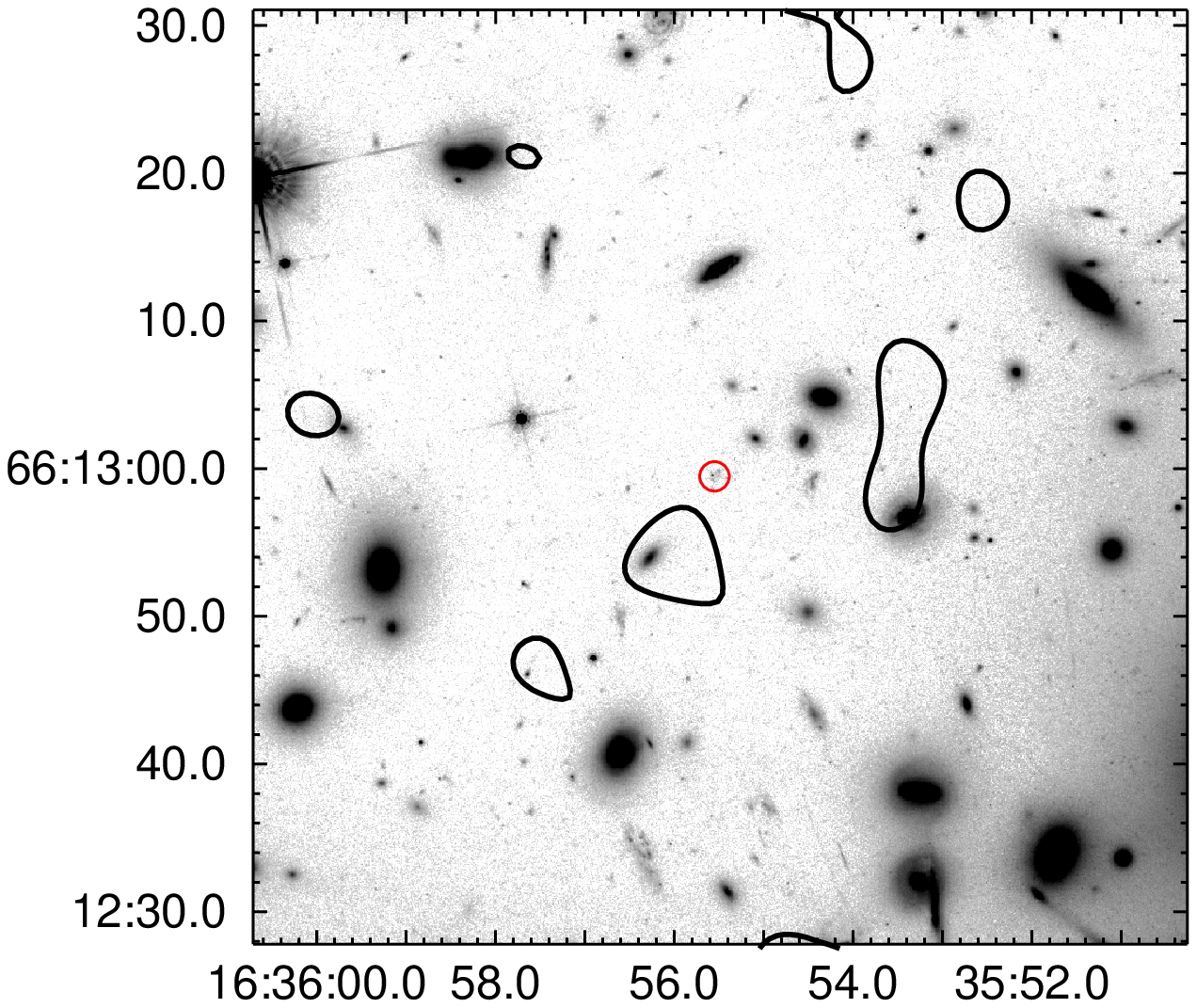}
\end{center}
\caption[]{The CO contours overlayed on the ACS F775W image.  Top:
SMMJ163555 with the contour levels 0.0005, 0.001, ... Jy/beam; middle:
SMMJ163541 with same contour levels, the red circle shows the optical
position; bottom:  SMMJ163556 with the contour level 0.00025 Jy/beam, the red
circle shows the optical position. 
\label{fig:optical}}
\end{figure}

Figure \ref{fig:res} shows the CO spectra and maps, and the resulting 
observational results are listed in Table \ref{tab:res}.  Additionally, the
CO contours have been overlayed on an optical image in Figure
\ref{fig:optical}. 
The CO(2-1) line for SMMJ163555 and the CO(3-2) line for SMMJ163541 were
successfully detected.
To obtain the fux, the line width, and the corresponding uncertainties,
which are given in Table \ref{tab:res}, we fit a single Gaussian to the
lines. 
The uncertainty of the integrated line intensity, $I_{\rm CO}$, is obtained
from the fitting of a single point source to the map.  The line intensity is
obtained from the Gaussian fit and is in good agreement with the integral
determined from the single point source fitting.  
We note that both lines are offset by -330 km/s and -400 km/s (bluewards)
from the UV (SMMJ163541) and optical (SMMJ163555) spectroscopic redshifts,
respectively.  
SMMJ163541 appears to have a second fainter CO component with a redshift of
$z=3.1883$, and while deeper observations are required to confirm this we
note that double-peak CO profiles appear to be common for SMGs
\citep{greve05}.  
The discrepancy between the optical and CO redshifts could be in either
case be due to a blend of two sources, in particular as we see in the case of
SMMJ163541.  There is no clear evidence for the presence of an AGN that could
drive an outflow.

CO(4-3) was undetected for SMMJ163556.  A $2.8\sigma$ continuum signal is
found at the position 16:35:56.01, +66:12:54.3; as this is $5''$ away from
the phase tracking center, we attribute this to noise; also no source was
found at this position at other wavelengths. 

The 3.0mm continuum emission was not detected for either source and we place
3$\sigma$ upper limits.  The fact that none of the sources were detected in
the continuum at 3mm is in agreement with the observed 850\,$\mu$m fluxes
\citep{knudsen06}.  
As the observations were carried out during the summer months, the conditions
were not useful for 1.3mm observations and neither line nor continuum was
detected.

%

\section{Discussion}
\label{sect:disc}

\subsection{Luminosities}

We derive the CO line luminosities from the velocity-integrated line
flux density following \citet{solomon97}.  
We estimate the far-infrared luminosity assuming a modified blackbody 
spectral energy distribution (SED) with a temperature of $T = 35$\,K and 
$\beta = 1.5$;  SMMJ163555 has a 450$\mu$m detection of 17.1\,mJy
\citep{knudsen08a}, thus the Rayleigh-Jeans tail is well-matched to this;
SMM163541 is detected in the 450$\mu$m map, however, it is very close to the
noisy edge of the map, thus the determined flux of $53\pm16$\,mJy might be
over-estimated, by possibly up to 50\%, and is thus consistent with a
temperature of 35\,K.  
For SMMJ163555 $L_{\rm FIR}$ is in good agreement with 
the ISOCAM result from \citet{biviano04}. 
The unlensed luminosities are given in Table~\ref{tab:prop}. 
The gravitational lensing magnification is determined using LENSTOOL
\citep{kneib93} using the detailed mass model of A2218 \citep{eliasdottir08}. 
It should be noted that the far-infrared luminosities of SMGs are often based
on only one or two submillimetre flux density data points and an assumed SED,
hence until more far-infrared wavelengths observations have been obtained for
SMGs, their far-infrared luminosities will be uncertain.  

\begin{figure*}
\includegraphics[width=8.5cm]{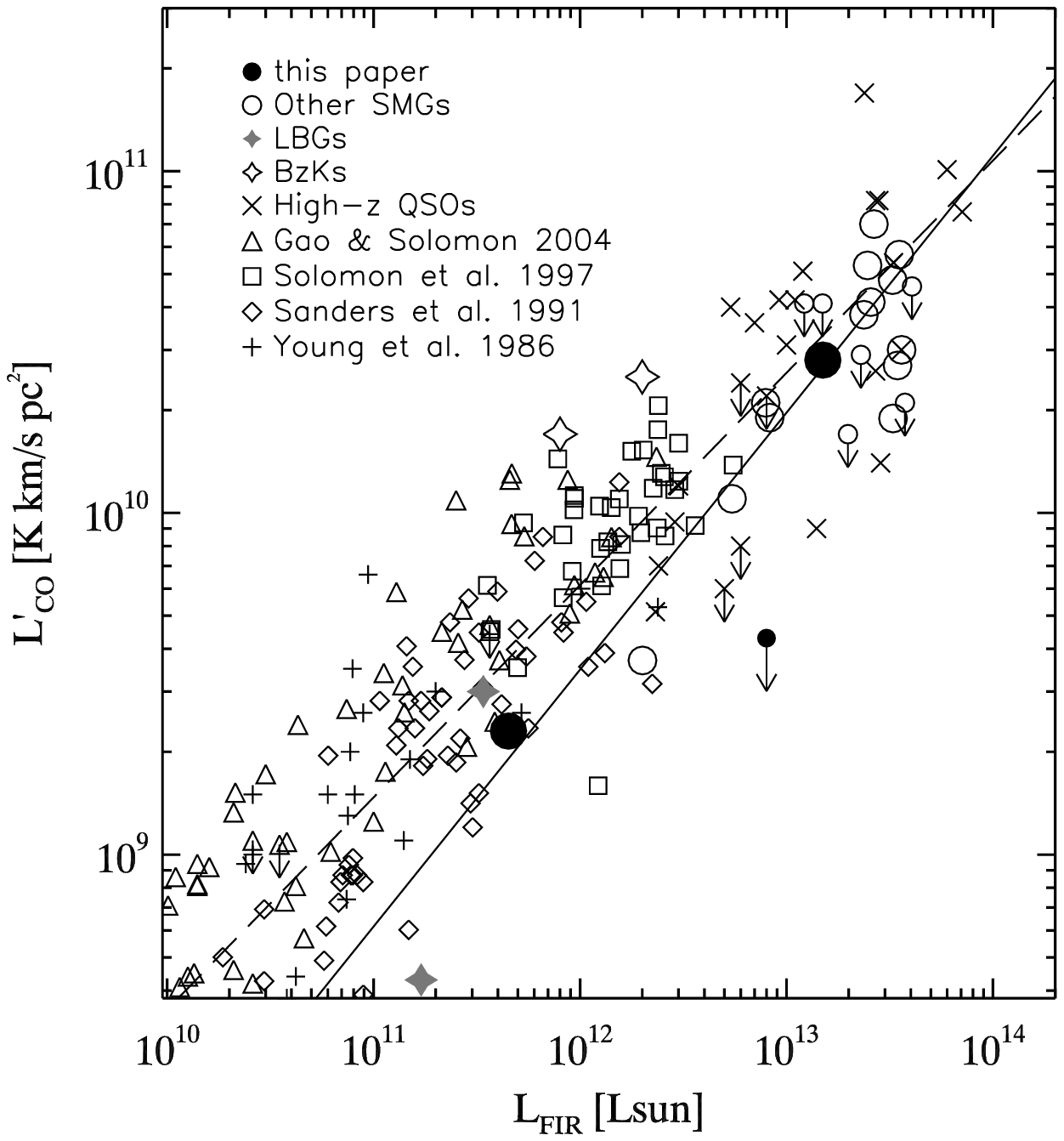}
\includegraphics[width=8.5cm]{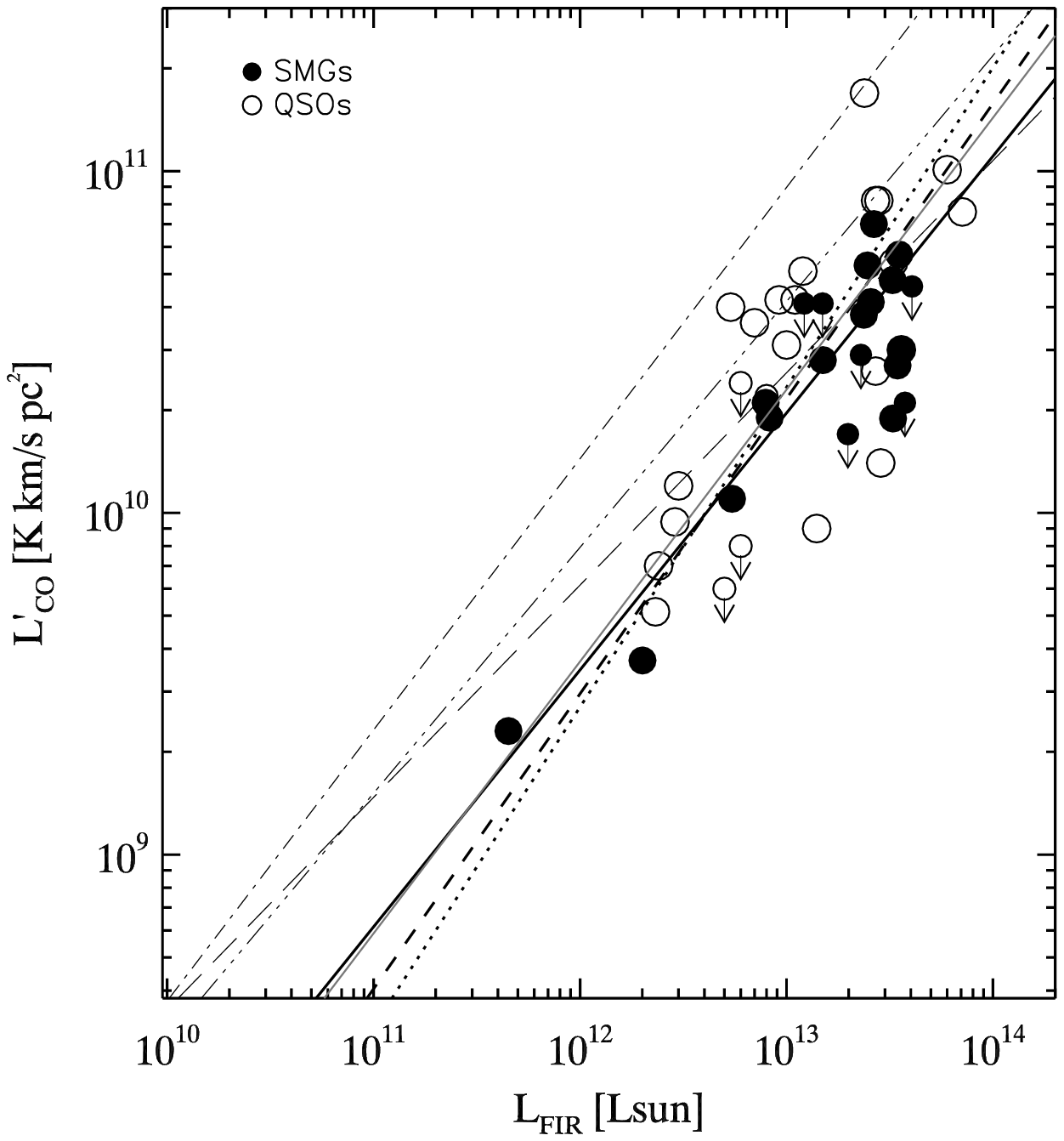}
\caption[]{Left panel: The CO luminosity plotted versus the far-infrared
luminosity for SMMJ163555 and SMMJ163541.  For comparison the results for other
submillimetre galaxies is plotted
\citep{frayer98,frayer99,neri03,genzel03,downes03,sheth04,kneib05,greve05,tacconi06}
along with low redshift and local ULIRG, LIRG and starburst galaxy results
\citep{young86,sanders91,solomon97,gao04}, high redshift quasars
\citep{barvainis94,omont96,guilloteau97,downes99,planesas99,guilloteau99,cox02,barvainis02,walter03,hainline04,riechers06,coppin08}
and LBG and BzK galaxies \citep{baker04,coppin07,daddi08}.
The solid line is the least-square fit to the SMG sample and the long-dashed line shows the best fit line to the $\log L_{\rm FIR} - \log
L{\rm '_{CO}}$ from \citet{greve05} for SMGs combined with low redshift
(U)LIRGs.
Right panel:  Same as above, though only plotting the high redshift SMGs and
QSOs.  The solid line is the least-square fit to the SMG sample, the dotted
line the fit to the QSO sample, the dashed line is the fit to the
combined high-$z$ sample, and the grey solid line is the fit to the combined
SMGs, QSOs, LBGs and BzKs sample.  The thin lines are from the previous fit to
combined high and low redshift samples, long-dashed from \citet{greve05} and
the dash-dot and dash-dot-dot-dot are from \citet{riechers06}, where the
former is the fit to the Gao \& Solomon sample and the latter is the fit to
the combined high redshift sources, the Gao \& Solomon sample, Solomon et al.
1997 sample, and PG QSOs samples.  
\label{fig:others}
}
\end{figure*}

It has been known for many years that a relation between the 
far-infrared luminosity and the CO luminosity of star forming galaxies
exists \citep[e.g.][]{young86,solomon97}.  This relation appears 
to be valid over a large range of luminosities, ranging from disk galaxies 
to high redshift QSOs.  
In Fig.~\ref{fig:others}, we compare the CO results with those from 
previous works and find that both SMMJ163555 and SMMJ163541 exhibit a 
similar behaviour.  
The $3\sigma$ $L'_{\rm CO}$ upper limit for SMMJ163556 is below the $L_{\rm
FIR}/L'_{\rm CO}$ relation indicating that the CO line might be outside of
the redshift range probed by our observations.  
We notice that the ratio $L_{\rm FIR}/L'_{\rm CO}$ of 350 for SMMJ163555  
is similar to that of SMMJ16359+6612 \citep{sheth04,kneib05}. 
These are both faint SMGs with unlensed submillimetre fluxes below the blank
field confusion limit of SCUBA at 850\,$\mu$m and have far-infrared
luminosities comparable to that of the local ULIRGs
\citep[e.g.][]{solomon97}, however their $L_{\rm FIR}/L'_{\rm CO}$ ratio is
higher than most of the ULIRGs in the \citet{solomon97} sample.  
A CO detection for other such faint SMGs is necessary 
to conclude whether this is a trend or just a consequence of the scatter.

We perform a straight-line least squares fit to the far-infrared luminosity
and the CO line luminosity of SMMJ163541, SMMJ163555, and the known high
redshift SMGs and QSOs
\citep{frayer98,frayer99,neri03,genzel03,downes03,sheth04,kneib05,greve05,tacconi06,barvainis94,omont96,guilloteau97,downes99,planesas99,guilloteau99,cox02,barvainis02,walter03,hainline04,riechers06,coppin08},
treating both $L_{\rm FIR}$ and $L'_{\rm CO}$ as independent variables with
uncorrelated errors and ignoring the upper limits.  For the combined SMGs and
QSO sample we find 
$\log L_{\rm FIR} = (1.16\pm0.05)\log L'_{\rm CO} +1.03$, 
for the SMG sample 
$\log L_{\rm FIR} = (1.33\pm0.07)\log L'_{\rm CO} -0.69$, 
and for the QSO sample 
$\log L_{\rm FIR} = (1.07\pm0.06)\log L'_{\rm CO} +1.92$. 
Similar results are obtained when excluding SMMJ163555. 
 When including the four other high redshift CO detections of LBG and BzK 
galaxies \citep{baker04,coppin07,daddi08} to the combined SMGs and QSO
sample, we find a consistent result: 
$\log L_{\rm FIR} = (1.26\pm0.04)\log L'_{\rm CO} -0.02$. 
In the luminosity range $L_{FIR} = 10^{11}-10^{14}$, the resulting three fits
are virtually identical, and the deviation from previous fits that combine
high and low redshifts samples \citep{greve05,riechers06} is no more than
$1\sigma$.  

With the CO detection of SMMJ163555 we can now probe the high redshift
luminous IR galaxy (LIRG) regime, which will otherwise only be accessible
with future instruments such as ALMA.  For the CO line luminosity of
SMMJ163555 of $2\times 10^{9}$\,K\,km/s\,pc$^{-2}$ the predictions for
$L_{\rm FIR}$, and hence the star formation rate (SFR), are a factor 3-5
higher for the high redshift galaxies compared to that of the nearby
galaxies, using the fits from above.  Given the rather sparse statistics for
the high redshift galaxies in the LIRG regime we cannot draw conclusions, but
we can speculate that the high redshift galaxies might have a higher $L_{\rm
FIR}$-to-$L'_{\rm CO}$ ratio, which can be used as a proxy for star formation
efficiency, than the nearby ones.   
{\citet{bouche07} also found in their study comparing the star formation
properties of various $z\sim2$ star forming galaxy populations that these
have higher star formation efficiency.}

\subsection{H$_2$ gas mass}

Assuming that the CO is thermalised to at least the J=3$\to$2 transition, 
we can convert the CO line luminosity into a molecular gas mass.  
We use a factor of $\alpha_{\rm CO} = M_{\rm H_2} / L'_{\rm CO(1-2)} = 
0.8 M_\odot / ({\rm K\,km/s\,pc^2}) = 0.2\alpha_{\rm CO}({\rm Galactic})$, 
as derived from observations of $z\sim 0.1$ ULIRGs \citep{downes98}.
The gas masses are probably uncertain by a factor of at least 2. 
For SMMJ163555, which is in the LIRG regime, the conversion factor may
well be closer to that for normal galaxies, hence the H$_2$ mass can be
correspondly higher.
The results, which have been corrected for gravitational lensing, are 
shown in Table \ref{tab:prop}. 

The line widths of SMMJ163555 and SMMJ163541 are 410 and 284 km/s, 
respectively, which is about half of the median value of 780 km/s
from the large sample studied by \citet{greve05}, but comparable to 
the values for the SMMJ16359+6612 \citep{kneib05}.  

The line profile can be used for deriving the line width, which we 
use for estimating a dynamical mass.  This estimate has large 
uncertainties due to the moderate signal to noise ratio of the spectrum, 
and the lack of spatial information due to the large beam. 
Assuming a disc model, we use 
$M_{\rm dyn}\sin ^{2}i\ (M_\odot)=4.2\times10^4\Delta v^2_{\rm FWHM}R$ 
\citep[see][]{neri03} 
to estimate the dynamical mass. 
First we solve the empirical Schmidt-Kennicutt relation as deduced for
high redshift starburst galaxies $\Sigma_{\rm sf}[{\rm M_\odot yr^{-1}
kpc^{-2}}] = 9.3\times10^{-5} ( \Sigma_{\rm gas} [{\rm M _{\odot}
pc^{-2}}])^{1.71}$ from \citet{bouche07} for the source size, deriving the
gas surface density from  $\sigma_{\rm gas} = f_{\rm gas}M_{\rm dyn}R/(\pi
R^2)$, assuming $f_{\rm gas} = 0.4$ and the star formation rate surface
density from $\Sigma_{\rm sf} = SFR / (\pi R^2)$.  Not knowing the
inclination, $i$, we apply an average $\langle \sin (i)\rangle = \pi/4$,
which means a correction of $\langle \sin ^2 (i) \rangle = 1.6$ for the
dynamical mass \citep[see e.g.][]{tacconi08}.   For SMMJ163555 and SMMJ163541
we estimate the source size of 0.49 kpc and 0.89 kpc. This corresponds to the
half-light radius, and to be able to compare with the results for other SMG
results from \citet{tacconi08} we estimate the dynamical mass within $2R$ and
find $M_{\rm dyn} = 1.1\times10^{10}$\,M$_\odot$ and
$1.0\times10^{10}$\,M$_\odot$, respectively.  
We note that for SMMJ163555, that despite the large uncertainties in this
estimate, this indicates that the molecular gas reservoir is a substantial
fraction of the total mass for these galaxies.  For SMMJ63541 we note that
the molecular gas mass is larger than the estimated dynamical mass, and we
assign this to limited knowledge on both $X_{\rm CO}$ and the dynamical state
of the SMMJ163541 (e.g.~the tentative second component would also contribute
and could indicate that SMMJ163541 are interacting galaxies).

\subsection{Star formation rate and efficiency}

The star formation efficiency (SFE), i.e.\ how efficiently the molecular gas
is being turned into stars, is often measured by the ratio $L_{\rm
FIR}/M_{\rm H_2}$.  We find a SFE of 
440\,L$_\odot$/M$_\odot$ for SMMJ163555 and 1000\,L$_\odot$/M$_\odot$ 
for SMMJ163541 indicating that both these galaxies are forming stars more 
efficiently than the average local ULIRGs \citep[e.g.][]{solomon97}. 
This is in good agreement with the previous SMG 
results from \citet{neri03} and \citet{greve05} who find a median 
value of $380\pm170$\,L$_\odot$/M$_\odot$ and 
$450\pm170$\,L$_\odot$/M$_\odot$, respectively.  

For a Salpeter Initial Mass Function with mass limits 0.1 and 100\,M$_\odot$ 
we estimate the star formation rate (SFR) through 
$1.755\times 10^{-10} \, L_{\rm FIR}/{\rm L_\odot}$ [M$_\odot$\,yr$^{-1}$] 
\citep{kennicutt98}.  For SMMJ163555 and SMMJ163541 we find an SFR of 80
M$_\odot$\,yr$^{-1}$ and 2600 M$_\odot$\,yr$^{-1}$.  
In rough numbers the gas depletion time-scale would be $\tau_{\rm SMG} \sim
20$ Myr and $\sim 8$ Myr for SMMJ163555 and SMMJ163541 respectively,  though
these numbers would be higher if the dust is partly heated by a central AGN.
The estimated time-scales are somewhat below that from \citet{greve05} of
$\tau_{\rm SMG} \sim 40$ Myr, however, the individual estimates are likely
affected by at least a factor two uncertainty arising from the assumptions
about the IMF, the conversion between CO line luminosity to H$_2$ mass, and
the far-infrared SED.  
In the case of SMMJ163555, which has a low far-infrared luminosity, it
is likely that the CO-H$_2$ conversion factor is larger by a factor 2-3,
implying a higher $M_{\rm gas}$ and therefore a time scale of 40-60 Myr.  For
SMMJ163541 we do not have a good constraint on the far-infrared SED and it is
likely that the deduced SFR is over-estimated, as demonstrated by
\citep{pope06} who showed that the 850$\mu$m-determined SFR of SMGs is larger
than that determined using 24$\mu$m and radio observations.  A lower SFR
would imply a higher depletion time-scale for SMMJ163541.  \citet{chapman08}
in a study of submm faint radio galaxies find a depletion time scale of
$\sim11$\,Myr, which compared with the stellar mass, could suggest that those
galaxies are seen towards the end of their current starburst episodes.
\citet{chapman08} also discuss that such small ages could imply a very high
duty cycle.


\section{Conclusions}
\label{sect:conc}

We have presented IRAM 3mm spectroscopy for three SMGs in the field of A2218.
CO detections were obtained for the sources SMMJ163555 at $z=1.0313$
(CO(2-1)) and SMMJ163541 $z=3.1824$ (CO(3-2)) confirming the optical
redshifts.  Both lines are shifted by $-400$ km/s relative to the optical
redshift.  The $z\approx 4$ candidate counterpart SMMJ163556 is undetected in 
CO(4-3) in the redshift range $4.035-4.082$.  
Future observations at high angular resolution and also other
J-transitions are required for a more detailed study of the SMGs.  Our
results add to the yet limited number of successful CO detections of SMGs. 

When correcting for gravitational lensing SMMJ163555 is so far the SMG known
with the lowest far-infrared luminosity and for the first time allows us to
probe the $L_{\rm FIR}-L'_{\rm CO}$ relation in the LIRG regime for high
redshift galaxies.  Based on this one detection we speculate that the high
redshift LIRGs have higher star formation efficiency than the nearby LIRGs.
This can, however, only be clarified with a large sample using future sensitive
instruments such as ALMA.

\begin{acknowledgements}
Based on observations carried out with the IRAM Plateau de Bure
Interferometer. IRAM is supported by INSU/CNRS (France), MPG (Germany) and
IGN (Spain).
This work has benefited from research funding from the European Community's
Sixth Framework Programme.  
JPK thanks CNRS for support.
This project was supported in part
by the Deutsche Forschungsgemeinschaft (DFG) via grant SFB 494 and 
by the DFG Priority Programme 1177. 
\end{acknowledgements}

\end{document}